\documentclass[preprint]{aastex}
\usepackage{graphicx}

\shorttitle{Pre-flare signatures}
\shortauthors{Kors\'os et al.}
\begin{document}

\title{On the evolution of pre-flare patterns of a 3-dimensional model of AR 11429}

\author{M. B. Kors\'os\altaffilmark{1,6}, S. Poedts\altaffilmark{4}, N. Gyenge\altaffilmark{1,6},  M. K. Georgoulis\altaffilmark{5}, S. Yu\altaffilmark{2}, S. K. Bisoi\altaffilmark{3}, Y. Yan\altaffilmark{3}, M. S. Ruderman\altaffilmark{1} \& R. Erd\'elyi\altaffilmark{1}}

\altaffiltext{1}{Solar Physics \& Space Plasma Research Center (SP2RC), School of Mathematics and Statistics, University of Sheffield, Hounsfield Road, S3 7RH, UK}
\altaffiltext{2}{New Jersey Institute of Technology, Center for Solar-Terrestrial Research, 323 Martin Luther King Blvd Newark, NJ 07102, United States}
\altaffiltext{3}{Key Laboratory of Solar Activity, National Astronomical Observatories, Chinese Academy of Sciences, Beijing, China}
\altaffiltext{4}{Center for Mathematical Plasma Astrophysics, KU Leuven, Celestijnenlaan 200B, 3001 Leuven, Belgium }
\altaffiltext{5}{Research Center for Astronomy and Applied Mathematics of the Academy of Athens, 4 Soranou Efesiou Street, 11527 Athens}
\altaffiltext{6}{DHO, Konkoly Astronomical Institute, Research Centre for Astronomy and Earth Sciences, Hungarian Academy of Sciences, Debrecen, P.O.Box 30, H-4010, Hungary }

\email{ korsos.marianna@csfk.mta.hu, robertus@sheffield.ac.uk, }

\begin{abstract}
We apply a novel pre-flare tracking of sunspot groups towards improving the estimation of flare onset time by focusing on the evolution of the 3D magnetic field construction of AR 11429. The 3D magnetic structure is based on potential field extrapolation encompassing a vertical range from the photosphere through the chromosphere and transition region into the low corona. The basis of our proxy measure of activity prediction is the so-called weighted horizontal gradient of magnetic field ($WG_{M}$) defined between spots of opposite polarities close to the polarity inversion line of an active region. The temporal variation of the distance of the barycenter of the  opposite polarities is also found to possess potentially important diagnostic information about the flare onset time estimation as function of height similar to its counterpart introduced initially in an application at the photosphere only in Korsos et al. (2015). We apply the photospheric pre-flare behavioural patterns of sunspot groups to the evolution of their associated 3D-constructed AR 11429 as function of height. We found that at a certain height in the lower solar atmosphere the onset time may be estimated much earlier than at the photosphere or at any other heights. Therefore, we present a tool and recipe that may potentially identify the optimum height for flare prognostic in the solar atmosphere allowing to improve our flare prediction capability and capacity.
\end{abstract}


\section{Introduction}
The 3D magnetic field structure of an AR is a key to obtain a more accurate information about the pre-flare evolution of a flaring AR. However, to construct an accurate 3D magnetic skeleton of an AR is a challenging task, see e.g. \cite{Wiegelmann12}. To accomplish this, one may need to measure the magnetic field at a number of heights. However, even nowadays, there are no routinely carried out direct measurements of the solar magnetic field in the solar atmosphere except only at the photosphere, see e.g. the magnetic field observations by SDO/HMI.
Measuring routinely the magnetic field higher up in the solar atmosphere proved to be far too challenging because of a number of technical reasons, see e.g. \cite{Solanki2003} and the follow-up debate by \cite{Judge2009}. Therefore, we may need to develop other complementary theoretical tools, that capture the key features of the 3D magnetic structure of an AR in the (lower) solar atmosphere. A frequently used such tool is magnetic field extrapolation based on photospheric measurements, see e.g. \cite{Wiegelmann12}. The method employs the approximation that the solar corona is considered to be in low-$\beta$ plasma state, or in other words, the magnetic pressure dominates over the pressure in the plasma. From this, we may approximate the plasma being in force-free (FF) state satisfying

\begin{equation}
{\bf j} \times {\bf B} =0.
 \label{force-free}
\end{equation}

Here, ${\bf j}=\frac{1}{\mu_{0}}\triangledown \times {\bf B}$ is the electric current density, {\bf B} the magnetic field and ${\mu_{0}}$ is the magnetic permeability of free space. This approximation may be a good one for describing the state of the coronal magnetic field. However, many studies claim (for good reasons) that the FF approximation is not reliable below the transition region, especially in the photosphere. 
Equation~(\ref{force-free}) can be satisfied by ${\bf j} = \triangledown \times {\bf B} =0$, where the current density vanishes everywhere. This is called a potential field (PF).
 Another option is when {\bf j} $||$ {\bf B}, that can be rewritten as $ \triangledown \times {\bf B} = \alpha \vec{B}$ and ${\bf B}\cdot  \triangledown \alpha =0$, where $\alpha$ is the FF parameter.  When $\alpha$  is constant along ${\bf B}$ everywhere in a given volume, it is the linear FF field (LFFF) approximation, otherwise  $\alpha$ may be a spatially dependent scalar function, in which case we have the nonlinear FF field (NLFFF) approximation.

It is elementary to understand the flare occurrence as the removal of free (i.e. non-potential) energy and helicity emerging through the Sun's surface, that is why the LFFF and NLFFF are more realistic (though not necessarily more pragmatic!) approximations than the PF extrapolation. Nevertheless, as a first step of our study, we use the PF extrapolation not only because of its mathematical simplicity but also to provide a first coarser view on how flare forecasting may be affected by tracking magnetic field structures in such 3D models of an AR in the lower solar atmosphere. Now, we only will investigate the typical pre-flare behaviour patterns as a function of height in the lower solar atmosphere.

\section{Analysis} 

\cite{Korsos2015}, hereafter K15, introduced a new parameter that may more accurately characterise the pre-flare evolution of the photospheric magnetic field of ARs. This is the weighted horizontal gradient of the magnetic field ($WG_{M}$) between nearby groups of spots of opposite magnetic polarities closest to the polarity inversion line.
With an empirical analyses, for all the observed flare cases between 1996-2010 that satisfy certain selection criteria (for that see K15), they discovered two typical pre-flare patterns for higher than M5 GOES 
flare-classes: 
(i) The pre-flare behavior of the $WG_{M}$ itself exhibits characteristic and unique patterns: steep rise, high maximum and a gradual decrease prior to flaring.
(ii) There is a U-shaped typical pre-flare pattern during the compressing and diverging motion in the distance value of the area-weighted barycenters of opposite polarities preceding the flare onset.  Also, K15 found, surprisingly, that if any one of these two typical pre-flare behaviours is missing then no flare occurs. 

Based on the empirical study of K15, a further tool was developed for estimating the flare onset time: if one can identify the time when the distance between the two barycentres of the area-weighted 
opposite polarities begins to grow again after the compression phase of the evolution, then one may be able to estimate the onset time of the flare by Eq. (3) of K15. With the aim to improve the efficiency of the onset time estimation capability of the $WG_{M}$ method that is based on the observed {\it photospheric} dynamics of pre-flare, we now propose to investigate the two pre-flare signatures not only at the photosphere but higher up in the lower solar atmosphere, i.e. in the Interface Region and low corona in a {\it 3D-constructed magnetic structure} for an AR. Implementing SDO/HMI line of sight magnetogram (and imaging) observations in the construction of the magnetic mapping of ARs may be a fruitful task. This procedure may also be cumbersome and computationally very expensive. Based on the $WG_{M}$ method, we test now the feasibility of estimating the flare onset time on the magnetic mapping of AR 11429 using one of the simplest 3D magnetic constructions, namelly the PF exploration (see Fig.~\ref{height}a). This way, if it works and applied routinely, a new data catalogue may be established, for further analysis. This catalogue which is based on the 3D magnetic data would contain valuable information on position, area and magnetic field/flux for each sunspot across the lower solar atmosphere.

In order to test the concept, here, let us now apply the $WG_{M}$ method and track the temporal variation of $WG_{M}$, distance of the area-weighted barycentre of the opposite polarities and the net flux at various heights for pre-flare dynamic studies for AR 11429 which produced one X5.4 (on 2012-03-07 00:24) and two M-class flares (M6.3 and M8.4 flares on 2012-03-09 03:53 and 2012-03-10 17:44). The investigation of the pre-flare dynamics is applied to the 3D construction of the AR between 6-12 March. Note, we cannot investigate the X5.4 flare because the limits of the $WG_{M}$ method (e.g. the X5.4 flare occurred close to the East limb).

\begin{figure}
\centering
\includegraphics[scale=0.74]{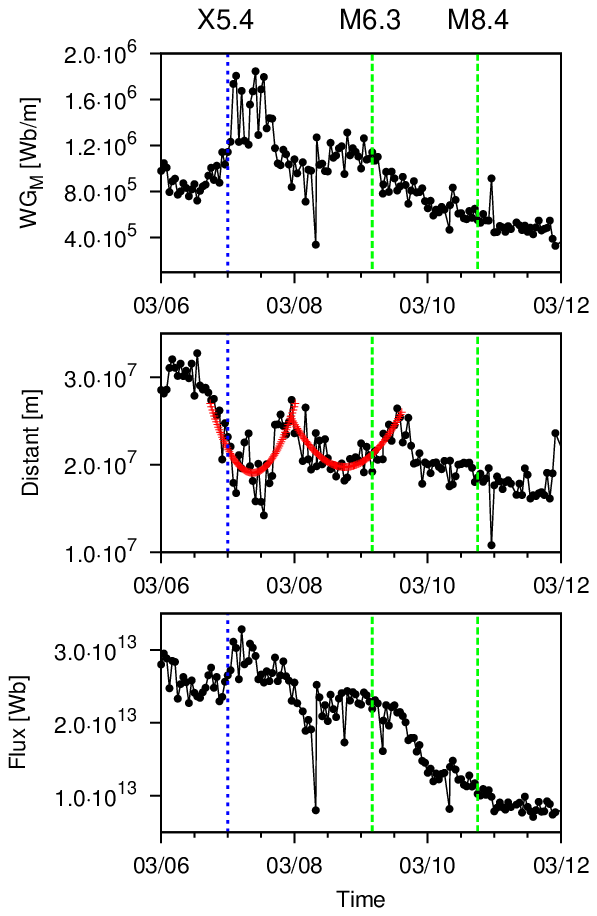}
\put(-130,185){(a)}
\includegraphics[scale=0.74]{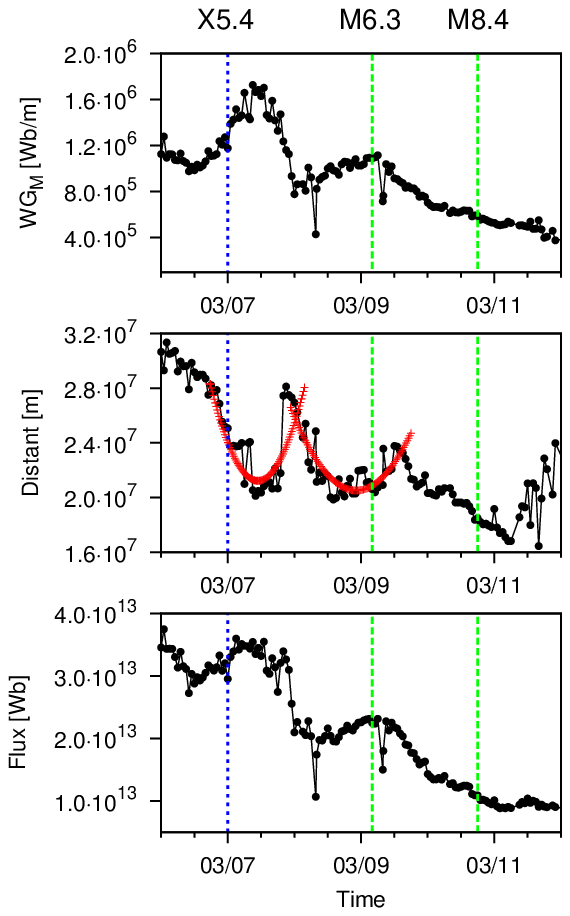}
\put(-130,185){(b)}
\includegraphics[scale=0.74]{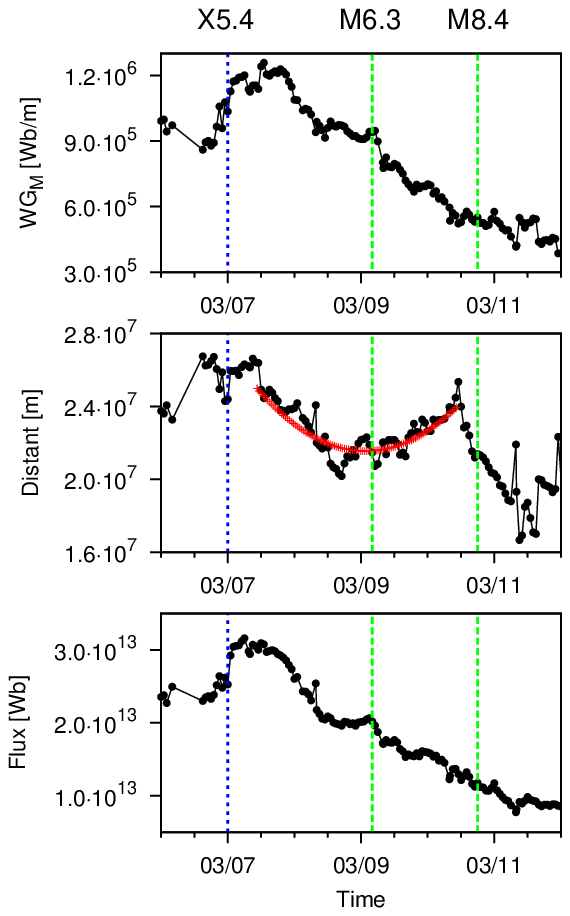}
\put(-130,185){(c)}

\caption{\label{fig11429}: Top panels are the variation of $WG_{M}$, middle panels depict the evolution of distance between the area-weighted barycenters of the spots of opposite polarities and the bottom panels show the unsigned 
flux of all spots at various heights as function of time. 
Column (a) is at photospheric level; (b) at 0.5 Mm and (c) at 1Mm levels, respectively, in the solar atmosphere. The red parabolae highlight the typical converging-diverging motion (U-shape) between the two area-weighted centres of the opposite polarities prior to flare(s). The blue/green vertical lines represent the X/two M-class flares onset times.}
\end{figure}

In the first step, as one ascends with a step-size of 100 km higher in the solar atmosphere and reaches the level of 2 Mm above the photosphere, following the application of the $WG_{M}$ method detailed in K15, we now recognise at most of the heights the prominent typical pre-flare behavior of the $WG_{M}$ (the increasing, maximum and after the decreasing phase of $WG_{M}$), and, the distance (evolving during the full converging-diverging motion of area-weighted barycenters) prior to flare(s). Furthermore, we also find that the evolution of $WG_M$ and distance changes remarkably as function of height (see Fig.~\ref{fig11429}a-c). 

In short, as we reach $\sim$0.4-0.6 Mm heights in the solar atmosphere, we see the beginning of the converging-diverging motion of the distance starts {\it earlier} than at the photospheric level.

\section{Result} 

\begin{figure} 
\centering
\includegraphics[scale=0.15]{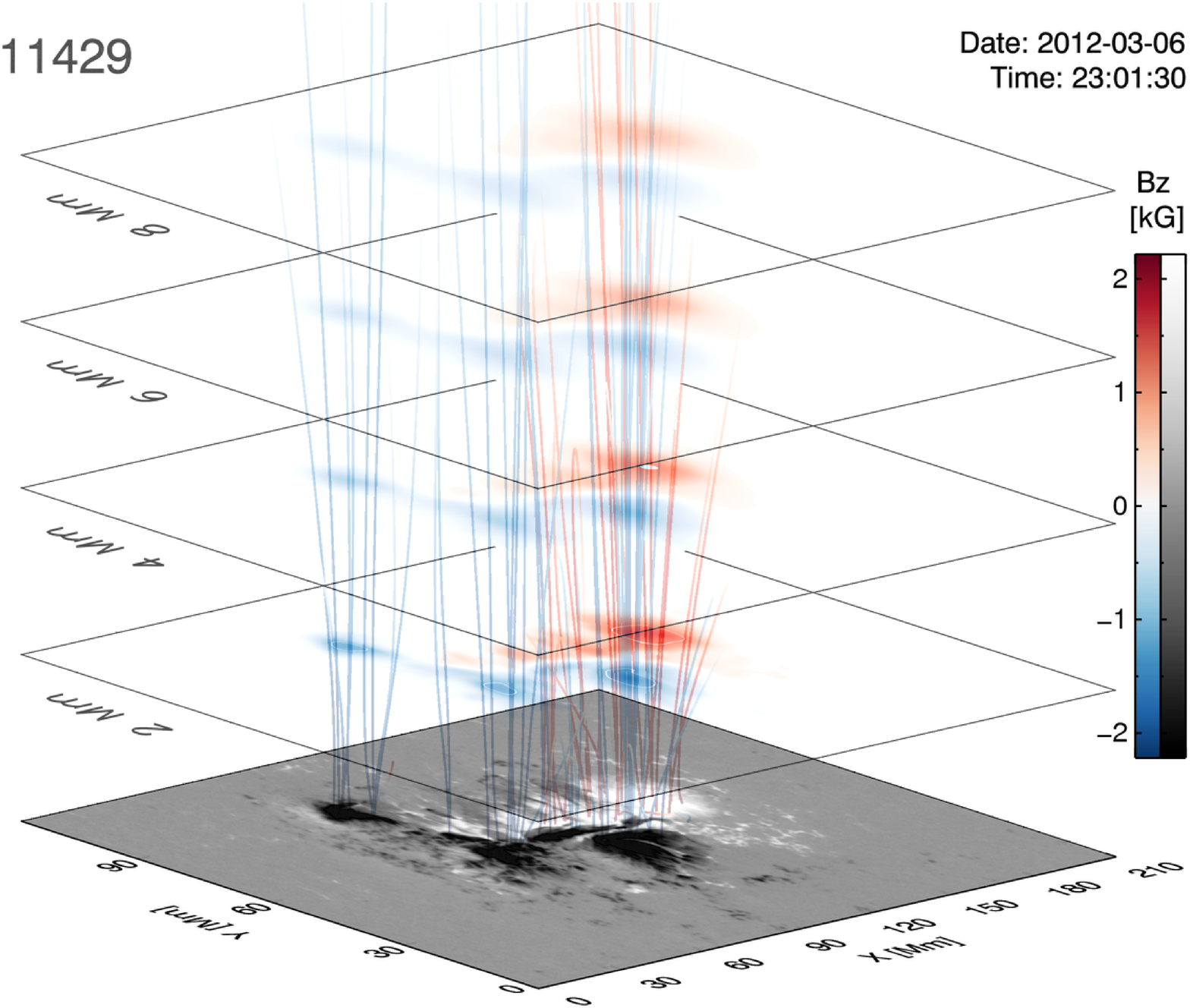}
\put(-193,140){(a)}
\includegraphics[scale=0.52]{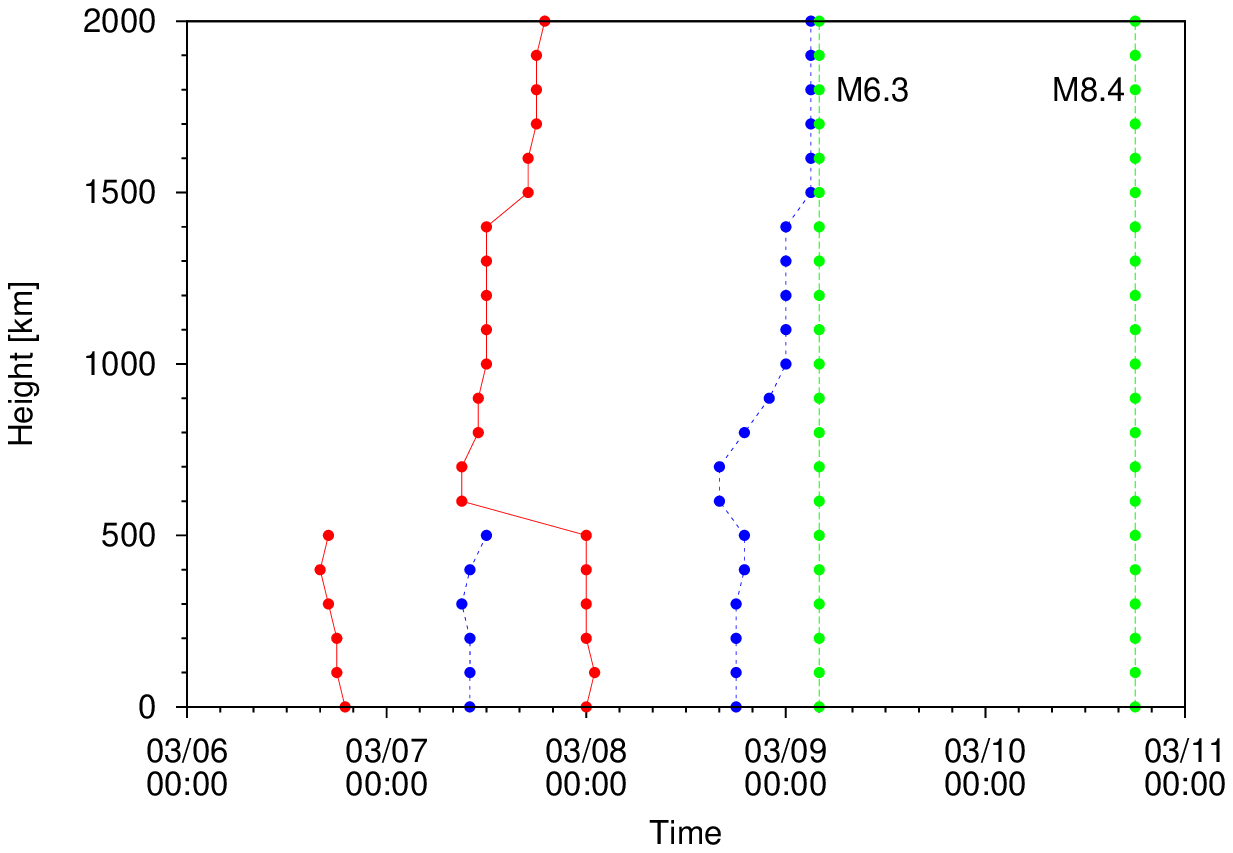} 
\put(-193,140){(b)}
\caption{(a): The 3D visualisation of AR 11429 is based on PF extrapolation from the photosphere to 8 Mm in the solar atmosphere; colour bars represent magnetic polarities (b) The green lines are associated with the onset time of M8.4 and M6.3-class flares, respectively. The red lines are the actual moment of start times of approaching (beginning of the red parabola in Fig.\ref{fig11429}), times of moments of the closest approaching point between two barycenters are the blue lines.}
\label{height}
\end{figure}

While applying the $WG_{M}$ method to the PF extrapolated structure of AR 11429 we notice that the starting time of the approaching phase {\it changes as function of height}, as visualized in Fig.~\ref{height}. Here, we plot the start of the converging phase (red lines) and the moment of time of the minimum distance of barycentres (blue lines) as function of height. We notice that in the case of the M6.3 flare at about 0.4 Mm level, and, in the case of M8.4 flare at about 0.6 Mm, the converging phase starts much earlier and reaches the minimum point also earlier than it does at the photosphere or at other heights. From this behaviour we may conclude that the flare onset times (by using the empirical Eq. 3 from K15 or otherwise) could be estimated sooner at certain heights than at photosphere or in the corona. 
In summary, identifying this optimum height in the solar atmosphere for a given AR may increase the time window elapsing prior to a flare. If this conclusion is confirmed by using a larger, statistically significant sample of ARs containing flare eruptions this would be a significant progress.

{\bf{\scriptsize Acknowledgement:} }
    {\tiny MBK, NGY, MSR and RE are grateful to the U. of Sheffield, STFC and the Royal Society (UK) for the support. YY is supported by NSFC grant.}

\end{document}